\documentclass[aps, prd,twocolumn,nofootinbib,tightenlines,floatfix,showpacs ,amssymb,superscriptaddress]{revtex4-2}
\usepackage{amsmath,graphicx}
\usepackage{mathrsfs}
\usepackage{subcaption}  

\begin{document}


\title{A Bayesian Framework for Constraining Magnetar Magnetic Fields from Repeating FRB Statistics}
\author{Can-Min Deng}
\email{dengcm@gxu.edu.cn}
\author{Hao-Hao  Chen}
\affiliation{Guangxi Key Laboratory for Relativistic Astrophysics, School of Physical Science and Technology, Guangxi University, Nanning 530004, People's Republic of China}


\begin{abstract}
		Fast radio bursts (FRBs) are widely considered to be associated with magnetars, motivated by the detection of an FRB-like radio burst from the Galactic magnetar SGR~1935+2154. However, constraining the magnetic field strength of extragalactic FRB sources remains challenging. In this work, we develop a Bayesian framework that models FRB time--energy sequences as a marked point process, combining burst waiting-time statistics with energy distributions to quantify the magnetic field strengths required to sustain the observed bursting activity under the magnetar powered scenario. Applying this method to a sample of repeating FRBs, we derive constraints on their magnetic fields by incorporating an empirical prior on the radio emission efficiency calibrated from the Galactic event. Under a conservative assumption for the activity duty cycle, most sources require magnetic energy reservoirs consistent with magnetar strength fields, with characteristic field strengths of order $10^{13}$--$10^{15}$ G, although the constraints remain sensitive to the poorly known efficiency and duty-cycle parameters. FRB~20200120E provides an interesting case with a substantially lower field requirement, highlighting the importance of source environment and evolutionary history in interpreting FRB activity. Our framework provides a statistical approach for connecting transient burst properties with magnetic energy reservoirs, with potential applications to FRBs and other magnetically powered transients.
\end{abstract}


\maketitle

\section{Introduction}
	Fast radio bursts (FRBs) are millisecond duration flashes of radio emission with brightness temperatures exceeding $10^{35}$ K \citep{Lorimer_2007, Thornton_2013}. Their large dispersion measures (DMs), together with precise interferometric localizations, have established their extragalactic origin and revealed host galaxies ranging from dwarf star forming systems to massive spirals \citep{Chatterjee_2017, Tendulkar_2017, Fong_2021, Ravi_2023}. Over the past decade, hundreds of FRBs have been discovered by wide-field facilities such as CHIME/FRB \citep{CHIMEFRB:2023myn}, revealing both apparently non-repeating events and repeating sources with diverse burst activity patterns \citep{Spitler_2016,2020ApJ...891L...6F,2021ApJ...923....1P,2022ApJ...926..206Z}. Despite these advances, the physical origin of FRB emission and the nature of their central engines remain open questions \citep{Zhang_2023}.

A major clue emerged in 2020 with the detection of an FRB-like radio burst from the Galactic magnetar SGR~1935+2154 \citep{CHIME_2020, Bochenek_2020}, accompanied by an X-ray burst \citep{Ridnaia_2021,2021NatAs...5..414K,2021NatAs...5..378L}. This event demonstrated that magnetars can produce bright coherent radio bursts and provided strong support for magnetars as viable progenitors of at least a fraction of FRBs. Magnetars are highly magnetized neutron stars with surface dipole fields typically in the range $10^{13}$--$10^{15}$ G \citep{2015RPPh...78k6901T,2017ARA&A..55..261K}, whose persistent and bursting emission is powered by the evolution and dissipation of magnetic energy \citep{1995MNRAS.275..255T,2006RPPh...69.2631H}. In theoretical models, crustal magnetic stresses generated by processes such as Hall drift can trigger fractures and magnetic reconnection events, providing a natural mechanism for releasing stored magnetic energy \citep{Pons_2007,2009ApJ...703.1044B}. Such magnetic energy dissipation has been widely considered as a possible power source for FRBs \citep{2014MNRAS.442L...9L,2017ApJ...843L..26B,2020ApJ...896..142B,2020MNRAS.498.1397L,2020ApJ...902L..32D,2020ApJ...904L..15I,2020ApJ...899..109W,Yang_2021}.

Determining the magnetic field strengths of extragalactic FRB progenitors, however, remains challenging. For Galactic magnetars, magnetic fields can be estimated from spin-down measurements or cyclotron absorption features \citep{2015RPPh...78k6901T,2017ARA&A..55..261K}, although these methods are themselves subject to model assumptions and primarily probe the external dipole component. For extragalactic FRBs, the lack of secure spin measurements and spectral diagnostics prevents direct field estimates. Previous attempts have mainly relied on either requirements imposed by specific radiation mechanisms, which inevitably depend on the adopted emission models \citep{2025ApJ...984...53Y}, or comparisons between the cumulative burst energy output and the available magnetic energy reservoir \citep{zhang_2025}. A systematic statistical framework that exploits the full temporal and energetic information contained in repeating FRB activity has not yet been established.

Here we develop a Bayesian framework that connects the observed burst statistics of repeating FRBs with the magnetic energy budget of magnetar engines. By treating the burst sequence as a marked point process, our approach simultaneously incorporates the temporal distribution of burst occurrence and the statistical distribution of burst energies. Under the assumption that FRBs are powered by magnetic energy dissipation, this framework allows us to quantify the magnetic field strengths required to sustain the observed activity of repeating sources. Applying this method to 12 repeating FRBs, we derive conservative constraints on their magnetic field strengths under physically motivated assumptions for the energy conversion efficiency and activity duty cycle. We find that most sources require magnetic energy reservoirs consistent with magnetar strength fields, while the inferred constraints remain strongly dependent on the poorly known efficiency and duty cycle parameters. Beyond FRBs, this framework provides a general statistical approach for connecting transient activity patterns with the underlying magnetic energy reservoirs of highly magnetized compact objects.

\section{The data} \label{data}
Our analysis is based on the repeating FRB dataset compiled by \cite{Xujy_2023}, which provides burst arrival times and fluence measurements for a large number of sources. From this catalog, we apply a set of selection criteria to ensure reliable statistical analysis. Specifically, we require each source to have more than ten detected bursts to obtain statistically meaningful constraints. We further exclude bursts lacking either time or fluence information, as well as waiting times longer than half a day, since such intervals are dominated by observational cadence rather than intrinsic activity.

Applying these filters yields a final sample of 12 relatively active repeating FRBs, which are listed in Table \ref{tab:frb}. For each burst, the isotropic-equivalent energy is calculated as
\begin{equation}
	E_{\mathrm{iso}}=\frac{4 \pi D_{\mathrm{L}}^{2} }{1+z} F\cdot \Delta \nu, 
	\label{Eiso}
\end{equation}
where $D_{\mathrm{L}}$ is the luminosity distance, $F$ is the fluence, $\Delta \nu$ is the bandwidth, and $z$ is the redshift. 
Here no explicit beaming correction is applied. For individual bursts, the observed isotropic-equivalent energy is reduced relative to the true emitted radio energy by the beaming factor ($f_b$). Under the assumption that burst locations and beam orientations are statistically distributed over the neutron-star surface, this reduction is compensated by a corresponding increase in the number of bursts occurring outside the observer's line of sight. Therefore, the time-averaged energy budget inferred from the observed burst population remains approximately unchanged.

An alternative description is that FRBs may be emitted within a larger-scale fan beam with solid angle ($\Delta\Omega$), introducing an effective global beaming factor ($F_b=\Delta\Omega/4\pi$) \citep{Zhang_2023}. However, such a correction should be distinguished from the geometry of the magnetic energy reservoir. Although the radio emission may originate from localized regions, for example along open magnetic field lines, the magnetic energy powering the activity is expected to be stored and dissipated on global magnetospheric and crustal scales, as suggested by magnetar bursts and flares observed at other wavelengths \citep{2006RPPh...69.2631H,2015RPPh...78k6901T,2017ARA&A..55..261K}.

Therefore, applying a global radio beaming correction to the total magnetic energy budget would implicitly assume that the magnetic dissipation itself is confined to the radio emitting beam, which is not required by magnetar burst physics. In our energy dissipation framework, the relevant quantity is the fraction of the dissipated magnetic energy converted into observable radio emission, represented by the efficiency factor ($\eta$), rather than the radio emission geometry alone. We therefore do not apply an additional global beaming correction.


For FRBs without spectroscopic redshifts( FRB 20201130A, FRB 20230607A, and FRB 20240619D),  we estimate their $z$ from the dispersion measure (DM) using the DM–$z$ relation \citep{Zhang_2023}:
\begin{eqnarray} \label{eq:DM_igm}
	\left\langle \mathrm{DM}_{\mathrm{IGM}}(z) \right\rangle=\frac{3cH_0\Omega_{\mathrm{b}}f_{\mathrm{IGM}}}{8\pi G m_\mathrm{p}}
	\int_0^z \frac{\chi(z')(1+z')dz'}{E(z')}, 
\end{eqnarray}
with $E(z')=\sqrt{\Omega_\mathrm{m}(1+z')^3+\Omega_\Lambda}$. Here $\chi(z')=7/8$ is the mean number of free electrons per baryon, $f_{\mathrm{IGM}}\simeq0.83$ is the baryon fraction in the intergalactic medium, and we adopt $\Lambda$CDM cosmological parameters from \citet{PlanckCollaboration_2020}. The observed DM is decomposed as
\begin{equation} \label{eq:DM}
	\mathrm{DM} = \mathrm{DM_{MW}} + \mathrm{DM_{halo}} + \mathrm{DM_{IGM}} + \frac{\mathrm{DM_{host}}}{1+z},
\end{equation}
where $\mathrm{DM_{MW}}$ is estimated using the NE2001 model and $\mathrm{DM_{halo}}$ is fixed at a conservative value of $50\,\mathrm{pc\,cm^{-3}}$ \citep{Cordes_2004}. $\mathrm{DM_{host}}$ is the  dispersion measure of the host galaxy.
For FRBs without host galaxy redshift measurements, we infer their redshifts using the updated Macquart relation by \citep{2026RAA....26k5016C}.
The inferred redshifts are then used to calculate the corresponding luminosity distances and burst energies. Since only three sources require this procedure, the uncertainty introduced by the DM–redshift conversion has a limited impact on our sample-level constraints. The dominant uncertainty in our analysis remains the poorly constrained radio-to-magnetic energy conversion efficiency ($\eta$).

The corresponding luminosity distances are then obtained via
\begin{eqnarray}
	D_{\mathrm{L}}=(1+z)\frac{c}{H_0}\int_0^z \frac{dz'}{E(z')}.
\end{eqnarray}

The final 12 sources sample is summarized in Table \ref{tab:frb}. The top panel lists FRBs with spectroscopic host-galaxy redshifts, while the bottom panel lists those with DM-inferred redshifts. The monitoring baseline  $\tau_{\rm b}$  for the observation is defined as the time interval between the first and last detections included in our dataset.

\begin{table}[tb]
	\centering
	\caption{The sample of repeating FRBs}
	\begin{tabular}{lccc}
		\hline\hline
		Name               & Redshift & $\tau_{\rm b}$ (year) & Reference       \\
		\hline
		FRB 20121102A      & 0.193    & 10.38             & [1--14]          \\
		FRB 20180301A      & 0.3304   & 4.57              & [15]             \\
		FRB 20180814A      & 0.0781   & 2.06              & [16], [17]       \\
		FRB 20180916B      & 0.0337   & 5.90              & [19--27]          \\
		FRB 20190520B      & 0.241    & 2.64              & [28--29]          \\
		FRB 20200120E      & 0.0008   & 4.31              & [30]             \\
		FRB 20201124A      & 0.0979   & 0.84              & [18], [28--37]    \\
		FRB 20220912A      & 0.0771   & 1.32              & [38--39]          \\
		FRB 20240114A      & 0.13     & 0.67              & [41--42]          \\
		\hline
		FRB 20201130A      & 0.112    & 0.35              & [18]             \\
		FRB 20230607A      & 0.215    & 1.65              & [40]             \\
		FRB 20240619D      & 0.310    & 0.02              & [43]             \\
		\hline\hline
	\end{tabular}
	\label{tab:frb}
	\vspace{0.5em}
	\footnotesize{Notes. The sample is sourced from the dataset curated by \citet{Xujy_2023}. Redshifts for samples in the top panel are measured through host galaxy identification, while those in the bottom panel are derived using the DM-$z$ relation. The FRB duration is defined as the time interval between the first discovery and the last detection date in our sample.\\
		Reference: [1] \citep{Spitler_2016}; [2] \citep{Scholz_2016}; [3] \citep{Spitler_2014}; [4] \citep{Michilli_2018}; [5] \citep{Li_2021}; [6] \citep{Oostrum_2020}; [7] \citep{Caleb_2020}; [8] \citep{Cruces_2020}; [9] \citep{Hewitt_2022}; [10] \citep{Law_2017}; [11] \citep{Zhang_2018}; [12] \citep{Pearlman_2020}; [13] \citep{Hilmarsson_2021}; [14] \citep{Li_2025}; [15] \citep{Kumar_2023}; [16] \citep{CHIME_2019_Natur}; [17] \citep{Xu_2025}; [18] \citep{CHIMEFRB:2023myn}; [19] \citep{Marcote_2020}; [20] \citep{Chime_FrbCollaboration_2020}; [21] \citep{Bethapudi_2023}; [22] \citep{Marthi_2020}; [23] \citep{Sand_2022}; [24] \citep{Chawla_2020}; [25] \citep{Pilia_2020}; [26] \citep{Pleunis_2021}; [27] \citep{Pastor_2021}; [28] \citep{Reshma_2023}; [29] \citep{Niu_2022}; [30] \citep{Nimmo_2023}; [31] \citep{Xu_2022}; [32] \citep{Zhang_2022}; [33] \citep{Hilmarsson_2021b}; [34] \citep{Marthi_2021}; [35] \citep{Kumar_2022}; [36] \citep{Lanman_2022}; [37] \citep{Nimmo_2022}; [38] \citep{Zhangyk_2023}; [39] \citep{Konijn_2024}; [40] \citep{Zhou_2025}; [41] \citep{zhang_2025}; [42] \citep{Tian_2024}; [43] \citep{Tian_2025}.}
\end{table}

\section{The method} \label{method}

\subsection{Linking the Magnetic Energy Dissipation to FRB Burst Statistics}
Magnetic energy stored in neutron stars provides a natural reservoir for magnetar burst activity across the electromagnetic spectrum. In magnetar crusts, magnetic evolution is regulated by several transport processes, including Hall drift, Ohmic decay, and ambipolar diffusion. Among these processes, Hall drift is expected to play an important role in redistributing magnetic stresses and transferring magnetic energy across spatial scales  \citep{Pons_2007}. Although the Hall effect is formally non-dissipative, it continually advects magnetic flux with the electron fluid and transfers energy from large-scale configurations to increasingly smaller structures. This cascade amplifies localized magnetic stresses until they exceed the crust’s elastic yield strength, leading to fractures and sudden magnetic reconnection events \citep{1995MNRAS.275..255T}. The associated magnetic energy release is thought to power the high energy outbursts observed from Galactic magnetars, and may likewise underlie FRBs as coherent emission episodes \citep{2020MNRAS.494.2385K,2020MNRAS.498.1397L,2021Univ....7...56L}.

On long timescales, the Hall drift also regulates the average magnetic energy dissipation rate in the magnetar crust. Although individual burst episodes are sporadic, the gradual Hall-driven evolution continuously converts magnetic energy over the timescale of $\tau_{\mathrm{Hall}}$  \citep{2006RPPh...69.2631H}. The corresponding long-term dissipation power can be expressed as
\begin{eqnarray}
	\langle L\rangle \simeq \frac{E_B}{\tau_{\text{Hall}}} = \frac{B^2 R_{\mathrm{NS}}^3}{6\tau_{\text{Hall}}},
	\label{eq:B_decay}
\end{eqnarray}
where $E_B=\frac{B^2}{8\pi}\left(\frac{4\pi}{3}R_{\mathrm{NS}}^3\right)$ is the magnetic energy reservoir and $R_{\mathrm{NS}}=10^6$ cm is the neutron-star radius. The Hall timescale is given by $\tau_{\mathrm{Hall}} = 1/(AB)$, where $A=10^{-19} ~\mathrm{G^{-1}yr^{-1}}$ \citep{Pons_2007} characterizes the crustal conductivity and electron density.
This steady dissipation provides the ultimate energy budget available to power magnetar-driven phenomena, including FRBs. If a fraction $\eta$ of the dissipated magnetic power is converted into coherent radio emission, then the average event rate of FRB bursts produced over the Hall timescale can be written as
\begin{eqnarray}
	\mathcal{R}_{\rm the}= \frac{\eta \langle L\rangle}{\langle E \rangle },
\end{eqnarray}
where $\langle E\rangle$ denotes the mean burst energy of a given source.

The above expression describes the long-term burst rate averaged over the Hall timescale, but observational monitoring typically spans only years—many orders of magnitude shorter than $\tau_{\mathrm{Hall}}$. Consequently, the instantaneous rate inferred from data may substantially exceed the long-term average, depending on how frequently the magnetar enters active bursting states. 
To account for the mismatch between the long-term magnetic evolution timescale and the observational monitoring duration, we introduce an effective activity fraction $D$.
The observed event rate can then be expressed as
\begin{eqnarray}
	\mathcal{R}= \frac{\eta \langle L\rangle}{D \langle E \rangle }= \frac{A \eta B^3 R_{\mathrm{NS}}^3}{6 D \langle E \rangle },
	\label{robs1}
\end{eqnarray}
where $D<1$ naturally accounts for the enhanced burst rate during active intervals relative to the long-term mean.
This formulation establishes a quantitative bridge between the internal magnetic energy dissipation of magnetars and the statistical properties of observed FRB bursts. In principle, Eq.~(\ref{robs1}) enables direct inference of the magnetic field strength if both $\eta$ and $D$ are independently known.

In practice, however, neither theoretical models nor existing observational data are currently sufficient to determine the duty cycle $D$ of magnetar burst activity. Consequently, within the present framework, $D$ must be treated as a free parameter. Likewise, the energy conversion efficiency $\eta$ remains poorly constrained, leading to a parameter degeneracy among $B$, $\eta$, and $D$ in Eq.~(\ref{robs1}).
To alleviate this issue, we employ a Bayesian approach in which $\eta$ is assigned an observational prior based on the Galactic FRB 20200428D (see section 3.2), while $D$ is conservatively fixed to its minimal physically allowed value $D_{\rm min}$.
Because the observed monitoring interval provides direct evidence that the source was active during this period, the most conservative choice is to associate this interval with the minimum possible active fraction, then we have $D_{\rm min} = \tau_{\rm b}/\tau_{\rm Hall}$, where $\tau_{\rm b}$ is the monitoring baseline of the FRB source. 
Thus, although the true value of $D$ remains uncertain, it must be at least as large as the ratio of the monitoring baseline to the Hall timescale, because the observed baseline is itself embedded within the long-term bursting activity over the full Hall timescale $\tau_{\rm Hall}$.
By adopting $D =D_{\rm min} = \tau_{\rm b}/\tau_{\rm Hall}$, Eq. (\ref{robs1}) reduces to
\begin{eqnarray}
	\mathcal{R}= \frac{\eta E_B}{\tau_{\rm b} \langle E \rangle }=\frac{\eta B^2 R_{\mathrm{NS}}^3}{6 \tau_{\rm b} \langle E \rangle },
	\label{robs2}
\end{eqnarray}
which corresponds to the energy-limited formulation adopted by \cite{zhang_2025}, now recovered as a special case of our general framework.
This treatment implies that, 
under this limiting assumption, the inferred magnetic fields represent conservative lower bounds within the adopted magnetic energy model.
Future observational constraints on $\eta$ and theoretical insights into the long-term activity duty cycle $D$ will be essential to fully break the $B$–$\eta$–$D$ degeneracy, enabling Eq.~(\ref{robs1}) to serve as a direct diagnostic of magnetar magnetic field strengths.

	\subsection{Bayesian Inference within a Marked Point Process Framework}
Constraining the magnetic field of magnetars with the framework outlined above requires observational access to the full burst sequence, i.e. the energy and occurrence time of each event, $\{(E_i, t_i)\}$. Mathematically, such data correspond to a \textit{marked point process}, in which the occurrence times $\{t_i\}$ form a discrete point process and each event is associated with a ``mark'' $\{E_i\}$ representing the burst energy \citep{Daley_2013}. This framework naturally allows us to characterize both the temporal statistics (through the distribution of waiting times) and the energy distribution of bursts, thereby providing the necessary basis for connecting the observed FRB activity with magnetic energy dissipation in magnetars.

When the burst energy $E$ and the occurrence time $t$ are assumed to be statistically independent, the marked point process can be factorized into two separate components. In this case, the temporal sequence $\{t_i\}$ is described by a probability distribution for waiting times (or event counts), while the energy sequence $\{E_i\}$ is represented by an independent probability distribution for burst energies. The joint probability of observing the sequence $\{(E_i,t_i)\}$ can then be written as the product of the two distributions, i.e.,
\begin{equation}
	P(E_i,t_i) = P(t_i) \times P(E_i),
\end{equation}
which provides a natural basis for modeling the temporal and energetic properties of FRBs separately while retaining the marked point process framework.

For the point process component, current observations of active repeating FRBs reveal clustering behavior, indicating that their burst activity is inconsistent with a homogeneous Poisson process and is better described by a more general Weibull renewal process \citep{Oppermann_2018,Zhang_2021, Tian_2025},
\begin{align} \label{weibull}
	P(\delta\mid k,\mathcal{R}) 
	& = k\delta^{-1} \left[ \delta \mathcal{R} \Gamma\left(1+\frac{1}{k}\right) \right]^{k} \nonumber \\
	& \quad \times \exp\left( - \left[ \delta \mathcal{R} \Gamma\left(1+\frac{1}{k}\right) \right]^{k} \right),
\end{align}
where $\Gamma$ is the gamma function and $\delta = t_i - t_{i-1}$ is the waiting time between bursts. Here, $k$ and $r$ are two free parameters representing the distribution shape and the average event rate, respectively. 
In practice, observational cadence and sensitivity can affect the observed waiting time distribution. We therefore restrict our analysis to burst intervals for which temporal coverage is sufficiently characterized and interpret the inferred rate parameters as effective observational rates.
Due to the absence of complete temporal coverage in the observational data, the waiting time associated with the first burst was excluded from the Weibull analysis. 

For certain sources where detections are not consistently achieved across all observational sessions, the Poisson distribution was employed to model event counts,
\begin{equation}
	P(N) = \frac{e^{- \mathcal{R} T_{\mathrm{e}}} (\mathcal{R} T_{\mathrm{e}})^N}{N!},
\end{equation}
where $N$ is the total number of bursts within the effective monitoring time $T_{\mathrm{e}}$. In our sample, this treatment applies to three sources, FRB 20180301A ($T_{\mathrm{e}}=193$ hr), FRB 20180814A ($T_{\mathrm{e}}=23$ hr), and FRB 20201130A ($T_{\mathrm{e}}=76.2$ hr). In particular, FRB 20180301A exhibits $k \simeq 1$ \citep{Kumar_2023}, for which a Poisson description is likewise adequate. In this way, the temporal part of the marked point process is specified either by a Weibull or Poisson model depending on source behavior, and it is analyzed using the Poisson process in this work.

For the mark distribution, we assume that the burst energies $\{E_i\}$ follow a log-normal distribution:
\begin{equation}
	P( E_i) = \frac{1}{E_i \sigma_{E}\,\sqrt{2\pi}}
	\exp\left(-\frac{( \log E_i - \log\langle E \rangle)^2}{2\sigma_{E}^2}\right),
\end{equation}
where $\sigma_{E}$ is the standard deviation of the sample and $\langle E \rangle$ depends on the magnetar field strength $B$, the energy conversion efficiency $\eta$, and the average event rate $r$ through the magnetic dissipation framework discussed above. This log-normal prescription thus specifies the statistical distribution of the marks in the process.

Under the independence assumption, the marked point process likelihood is simply the product of the temporal and energetic components. The corresponding log-likelihood for the observed sequence $(E_i,t_i)$ is then
\begin{equation}
	\begin{split}
		\ln P (k,\mathcal{R},B,\eta|E_i,t_i) = 
		\sum_{i=2}^{n} \big[\ln P(k,\mathcal{R} |t_i)\, \\ +\,\ln P(B,\eta,\mathcal{R} |E_i)\big]. 
	\end{split}
\end{equation}
This formulation explicitly casts the FRB sequence as a marked point process, with the point component capturing burst occurrence statistics and the mark component describing the energy distribution. It therefore provides a principled statistical connection between the observed FRB activity and the underlying magnetar magnetic energy dissipation, enabling quantitative inference of their magnetic field strengths.

\begin{table*}[tb]
	\centering
	\caption{Observational Parameters of the FRB and X‑ray from SGR 1935+2154 on 28 April 2020}
	\begin{tabular}{lcccc}
		\hline\hline
		Parameter & \multicolumn{2}{c}{\textbf{CHIME}} & \multicolumn{2}{c}{\textbf{Konus‑Wind}} \\
		\cline{2-3} \cline{4-5}
		& Component 1 & Component 2 & Component 1 & Component 2 \\
		\hline
		Arrival time, 28 April 2020 (UTC, geocentric)
		& 14:34:24.42650 & 14:34:24.45547 & 14:34:24.427 & 14:34:24.455 \\
		Fluence ($\mathrm{kJy\,ms}$)
		& 480 & 220 & --- & --- \\
		Flux ($\times 10^{-6}$ $\mathrm{erg\,cm^{-2}\,s^{-1}}$)
		& --- & --- & $7.5 \pm 1.2$ & $9.1 \pm 1.5$ \\
		Energy ($\mathrm{erg}$)
		& $1.86 \times 10^{34}$ & $8.53 \times 10^{33}$ & $2.91 \times 10^{38}$ & $3.53 \times 10^{38}$ \\
		\hline\hline
	\end{tabular}
	\label{tab:FRB_X-ray}
	\vspace{0.4em}
	\footnotesize{Notes. Energy values assume a distance of $9\,\mathrm{kpc}$ to the source. X‑ray energies calculated with $4\,\mathrm{ms}$ pulse width \citep{Ridnaia_2021}.}
\end{table*}

	In the marked point process framework introduced above, the likelihood depends on the temporal parameters $(k,\mathcal{R})$, the magnetar field strength $B$, and the conversion efficiency $\eta$. To complete the Bayesian inference, prior distributions must be specified for these parameters. Owing to insufficient prior information, we adopt uniform priors in logarithmic space for $k$, $\mathcal{R}$, and $B$.  By contrast, the prior for $\eta$ is constructed hierarchically from observational data, due to $\eta$ and $B$ are strongly degenerate, both appear jointly in the energy scaling of FRBs.
Specifically, we model $y=\log \eta$ as a normally distributed variable,
\begin{equation}
	P(y) = \frac{1}{ \sqrt{2\pi} \sigma} \exp\left(-\frac{(y - \mu)^2}{2\sigma^2}\right),
\end{equation}
and update this distribution using the observed values $\{y_n\}$. The predictive distribution is then
\begin{eqnarray} \label{y_post}
	p(y|y_n) = \iint p(y|\mu, \sigma^2)\,p(\mu, \sigma^2|y_n)\,d\mu d\sigma^2 .
\end{eqnarray}

Integrating Eq. (\ref{y_post}) under non-informative hyper-parameter priors yields a posterior predictive distribution of the form
\begin{equation}
	P(y|y_n) \sim t_{n-1}\left(\bar{y}_n, s^2\left(1 + \frac{1}{n}\right)\right),
\end{equation}
where $\bar{y}_n$ is the sample mean, $s^2$ is the sample variance, and $n$ is the number of observations. 
This result highlights that, under non-informative priors, the Bayesian predictive distribution of a normal mean is a $t$-distribution, a classical result in Bayesian statistics \citep{Gelman_2013}.  As the sample size $n$ increases, this predictive distribution converges toward a Gaussian, thereby allowing progressively tighter constraints on $\eta$. The detailed derivation process is provided in the the Appendix A.

In this work, the prior for $\eta$ is constrained by the joint radio and X-ray detection of the Galactic magnetar SGR 1935+2154 on 28 April 2020, when Konus-Wind detected a double-peaked X-ray burst coincident with the bright double-peaked FRB-like radio burst observed by CHIME (see Table \ref{tab:FRB_X-ray}) \citep{CHIME_2020,Ridnaia_2021}. The radio-to-X-ray energy ratios from this event yield two estimates, $\eta_{\rm r/x,1} = 2.4 \times 10^{-5}$ and $\eta_{\rm r/x,2} = 6.4 \times 10^{-5}$, corresponding to $y_1 = -4.6$ and $y_2 = -4.2$, respectively. These measurements provide $n=2$, $\bar{y}_n=-4.4$, and $s=0.30$. As a conservative estimate, we adopt these values as the prior constraint on $\eta$ in our inference  \footnote{Strictly speaking, these ratios do not represent the intrinsic efficiency of converting magnetic dissipation energy into radio emission, since the total dissipated magnetic energy may exceed the observed X-ray output. Instead, they should be regarded as conservative upper limits on the radio conversion efficiency, i.e., $\eta\leq\eta_{\rm r/x}$. Therefore, adopting the observed radio-to-X-ray ratios as the prior constraint on $\eta$ provides the most conservative estimate: any lower intrinsic efficiency would require a larger magnetic energy reservoir and consequently lead to stronger constraints on the magnetar magnetic field strength.
}.

This treatment ensures that $\eta$ is not regarded as entirely unconstrained, but instead anchored to empirical Galactic observations. In doing so, it mitigates the degeneracy between $\eta$ and $B$, thereby allowing meaningful inference on the magnetic field strength of FRB progenitors. As more multi-wavelength FRB counterparts are detected, the prior on $\eta$ can be progressively refined, leading to tighter and more robust constraints on magnetar magnetic fields within this Bayesian framework.

\section{Results} \label{result}	
We employed Markov Chain Monte Carlo (MCMC) sampling to explore the posterior distributions of the model parameters within the Bayesian framework described above. To ensure a broad coverage of the parameter space, we initialized 40 independent chains with starting values drawn uniformly from the ranges $10^{12}$–$10^{17}$~G for the magnetic field $B$ and $10^{-8}$–$1$  for the efficiency factor $\eta$. Each chain was run for 100,000 iterations, and the procedure was repeated ten times to assess robustness. Throughout the analysis, we imposed $B>10^{12}$ G and $\eta<1$ as physical bounds. Two representative cases were considered: (i) a conservative duty cycle $D = D_{\rm min}=\tau_{\mathrm{b}}/\tau_{\mathrm{Hall}}$, and (ii) a fixed duty cycle $D=0.01$. The first case provides lower-limit constraints on $B$, while the second illustrates the effect of adopting a larger but still plausible active fraction.

Table~\ref{cons_result} summarizes the results for the conservative duty-cycle case. The inferred magnetic fields span approximately $B \gtrsim  10^{12} - 10^{15}$ G, with broad posterior intervals of up to three orders of magnitude. FRB 20200120E yields the lowest field strength, $B \gtrsim 2.5 \times 10^{12}$ G, consistent with its location in an old stellar population in the M81 globular cluster \citep{Kirsten_2022}. For most other sources, the posterior medians exceed $10^{14}$ G, with several reaching $> 10^{15}$ G, though the large uncertainties mean that values below $10^{14}$ G cannot yet be excluded. For instance, the ten-year dataset of FRB 20121102A implies $\log(B/{\rm G}) \gtrsim  15.4^{+1.3}_{-1.7}$, while FRB 20240114A gives $\log(B/{\rm G}) \gtrsim  15.1^{+1.5}_{-1.6}$, in agreement with \citet{zhang_2025}. The shape parameter $k$ further confirms that most FRBs exhibit clustered burst activity $k<1$, deviating from a homogeneous Poisson process. The broad credible intervals on $\eta$ reflect both its strong degeneracy with $B$ and the limited observational prior (based solely on the FRB 20200428D event), rendering the constraints intentionally conservative.

To test the sensitivity of our inference to the duty cycle, we repeated the analysis with $D=0.01$, corresponding to the assumption that the magnetar is active for only 1\% of the Hall timescale. The results, summarized in Table~\ref{ext_result}, show systematically larger field strengths, with typical posterior medians in the range $B \sim 10^{15}$–$10^{16}$~G. Only FRB~20200120E again stands out with a much lower field, $\log(B/{\rm G})=14.0^{+0.7}_{-0.5}$. These higher field estimates arise because, for a larger $D$, the observed burst activity must be powered by a correspondingly larger magnetic energy reservoir. At the same time, the credible intervals are narrower than in the conservative case, reflecting the fact that once $D$ is specified, the inference relies on Eq.~(\ref{robs1}), where $B$ enters cubically, rather than Eq.~(\ref{robs2}), where $B$ enters quadratically. This stronger dependence on $B$ naturally enhances the sensitivity of the constraints.

Taken together, these results demonstrate that our conservative assumption ($D=\tau_{\mathrm{b}}/\tau_{\mathrm{Hall}}$) yields robust lower limits on the magnetar field strengths of FRB progenitors, disfavouring weak-field ($<10^{13}$~G) scenarios
except in special cases such as FRB 20200120E. When plausible and larger  duty cycles are adopted, the required fields rise well into the magnetar regime ($\gtrsim10^{15}$~G). The current uncertainties in both $\eta$ and $D$ mean that our present framework provides conservative bounds rather than precise measurements. However, the methodology also highlights a clear path forward: if future multi-wavelength observations can better constrain $\eta$, and if long-term monitoring can shed light on the duty cycle $D$, then Eq.~(\ref{robs1}) could evolve from a lower-limit estimator into a direct and sensitive diagnostic of magnetar magnetic fields in FRBs.

\begin{table}[tb]
	\centering
	\caption{Parameter inference with duty cycle $D=\tau_{\mathrm{b}}/\tau_{\mathrm{Hall}}$.}
	\begin{tabular}{lcccc}
		\hline\hline
		Name            & $\log (B/\mathrm{G})$ & $\log (\mathcal{R}/\mathrm{day})$ & $\log \eta$ & $k$ \\
		\hline
		FRB 20121102A   & $15.4^{+1.3}_{-1.7}$ & $2.7^{+0.1}_{-0.1}$ & $-4.3^{+3.5}_{-2.5}$ & $0.6^{+0.1}_{-0.1}$ \\
		FRB 20240114A   & $15.1^{+1.5}_{-1.6}$ & $3.9^{+0.1}_{-0.1}$ & $-4.1^{+3.2}_{-2.9}$ & $0.5^{+0.1}_{-0.1}$ \\
		FRB 20180301A   & $15.0^{+0.9}_{-0.9}$ & $0.7^{+0.1}_{-0.1}$ & $-4.4^{+1.7}_{-1.7}$ & --- \\
		FRB 20190520B   & $15.0^{+0.9}_{-0.7}$ & $1.8^{+0.2}_{-0.3}$ & $-4.4^{+1.4}_{-1.8}$ & $0.4^{+0.1}_{-0.1}$ \\
		FRB 20230607A   & $15.0^{+1.6}_{-1.3}$ & $3.1^{+0.1}_{-0.1}$ & $-4.4^{+2.5}_{-3.1}$ & $0.8^{+0.1}_{-0.1}$ \\
		FRB 20240619D   & $15.0^{+1.0}_{-1.0}$ & $3.1^{+0.1}_{-0.1}$ & $-4.4^{+2.0}_{-1.9}$ & $0.7^{+0.1}_{-0.1}$ \\
		FRB 20220912A   & $14.9^{+1.2}_{-1.5}$ & $2.9^{+0.1}_{-0.1}$ & $-4.4^{+3.1}_{-2.5}$ & $0.5^{+0.1}_{-0.1}$ \\
		FRB 20201124A   & $14.7^{+1.5}_{-1.6}$ & $2.6^{+0.1}_{-0.1}$ & $-4.4^{+3.2}_{-3.0}$ & $0.4^{+0.1}_{-0.1}$ \\
		FRB 20180916B   & $14.6^{+1.0}_{-0.8}$ & $1.5^{+0.1}_{-0.1}$ & $-4.4^{+1.6}_{-2.0}$ & $0.8^{+0.1}_{-0.1}$ \\
		FRB 20180814A   & $14.4^{+1.0}_{-0.8}$ & $1.0^{+0.2}_{-0.2}$ & $-4.4^{+1.5}_{-2.0}$ & --- \\
		FRB 20201130A   & $14.1^{+0.9}_{-0.8}$ & $0.6^{+0.2}_{-0.2}$ & $-4.4^{+1.5}_{-1.7}$ & --- \\
		FRB 20200120E   & $12.9^{+1.0}_{-0.5}$ & $3.1^{+0.2}_{-0.2}$ & $-4.4^{+1.0}_{-1.9}$ & $0.6^{+0.1}_{-0.1}$ \\
		\hline\hline
	\end{tabular}
	\label{cons_result}
	\vspace{0.4em}
	\footnotesize{Notes. Posterior medians with 90\% credible intervals. The duty cycle $D$ is set to the most conservative lower bound $\tau_{\rm b}/\tau_{\rm Hall}$, so the derived magnetic fields represent robust \textit{lower limits}. For FRB 20180814A and FRB 20201130A, we adopt Poisson statistics due to incomplete burst detections. FRB 20180301A is likewise analyzed with a Poisson distribution since $k\simeq 1$ \citep{Kumar_2023}.}
\end{table}

\begin{table}[tb]
	\centering
	\caption{Parameter inference with fixed duty cycle $D=0.01$.}
	\begin{tabular}{lcccc}
		\hline\hline
		Name            & $\log (B/\mathrm{G})$ & $\log (\mathcal{R}/\mathrm{day})$ & $\log \eta$ & $k$ \\
		\hline
		FRB 20240619D   & $16.2^{+0.6}_{-1.0}$ & $3.1^{+0.1}_{-0.1}$ & $-4.4^{+2.9}_{-1.8}$ & $0.7^{+0.1}_{-0.1}$ \\
		FRB 20240114A   & $15.8^{+1.0}_{-1.2}$ & $3.9^{+0.1}_{-0.1}$ & $-4.0^{+3.5}_{-3.1}$ & $0.5^{+0.1}_{-0.1}$ \\
		FRB 20121102A   & $15.6^{+1.1}_{-1.2}$ & $2.7^{+0.1}_{-0.1}$ & $-4.2^{+3.6}_{-3.2}$ & $0.6^{+0.1}_{-0.1}$ \\
		FRB 20220912A   & $15.6^{+0.8}_{-1.1}$ & $2.9^{+0.1}_{-0.1}$ & $-4.4^{+3.2}_{-2.4}$ & $0.5^{+0.1}_{-0.1}$ \\
		FRB 20230607A   & $15.6^{+1.0}_{-1.1}$ & $3.1^{+0.1}_{-0.1}$ & $-4.3^{+3.3}_{-2.9}$ & $0.8^{+0.1}_{-0.1}$ \\
		FRB 20180301A   & $15.5^{+0.6}_{-0.6}$ & $0.7^{+0.1}_{-0.1}$ & $-4.4^{+1.7}_{-1.9}$ & --- \\
		FRB 20190520B   & $15.5^{+0.7}_{-0.5}$ & $1.8^{+0.2}_{-0.3}$ & $-4.4^{+1.5}_{-2.1}$ & $0.4^{+0.1}_{-0.1}$ \\
		FRB 20201124A   & $15.5^{+1.0}_{-1.2}$ & $2.6^{+0.1}_{-0.1}$ & $-4.2^{+3.5}_{-2.9}$ & $0.4^{+0.1}_{-0.1}$ \\
		FRB 20180814A   & $15.2^{+0.7}_{-0.5}$ & $1.0^{+0.2}_{-0.2}$ & $-4.4^{+1.5}_{-2.1}$ & --- \\
		FRB 20180916B   & $15.2^{+0.7}_{-0.5}$ & $1.5^{+0.1}_{-0.1}$ & $-4.4^{+1.6}_{-2.2}$ & $0.8^{+0.1}_{-0.1}$ \\
		FRB 20201130A   & $15.2^{+0.7}_{-0.5}$ & $0.6^{+0.2}_{-0.2}$ & $-4.4^{+1.6}_{-2.0}$ & --- \\
		FRB 20200120E   & $14.0^{+0.7}_{-0.5}$ & $3.1^{+0.2}_{-0.2}$ & $-4.4^{+1.6}_{-2.0}$ & $0.6^{+0.1}_{-0.1}$ \\
		\hline\hline
	\end{tabular}
	\label{ext_result}
	\vspace{0.4em}
	\footnotesize{Notes. Posterior medians with 90\% credible intervals. Here the duty cycle is fixed at $D=0.01$ to illustrate how tighter assumptions on activity impact the inferred parameters. Compared to the conservative case ($D=\tau_{\rm b}/\tau_{\rm Hall}$), this choice leads to systematically higher inferred magnetic fields and smaller uncertainties, since Eq.~(\ref{robs1}) depends more sensitively on $B$ (cubic scaling rather than quadratic).}
\end{table}

	\section{Discussion} \label{discussion}

In this work, we develop a Bayesian framework to constrain the magnetic fields of repeating FRB progenitors under the hypothesis that FRB activity is ultimately powered by magnetar magnetic energy dissipation. By treating observed FRB bursts as a marked point process, our method simultaneously incorporates the temporal clustering properties and burst energetics, providing a statistical connection between burst activity and the underlying magnetic energy reservoir. Applying this framework to 12 repeating FRBs, we obtain conservative constraints on their magnetic field strengths. For most sources, the inferred fields are within the magnetar regime ($\gtrsim10^{13}$~G), while FRB~20200120E remains consistent with a lower-field solution ($\gtrsim10^{12}$~G)
, in agreement with its origin in an old globular cluster environment.

A key aspect of our analysis is the treatment of the duty cycle $D$. Because the long-term active fraction of magnetar bursting activity is currently unknown, we adopt the minimal physically allowed value, $D=\tau_{\rm b}/\tau_{\rm Hall}$, where $\tau_{\rm b}$ is the observed monitoring baseline. This corresponds to the most conservative scenario in which the observed bursting phase represents only a small fraction of the Hall evolution timescale. Under this assumption, the inferred magnetic fields represent robust lower limits. Therefore, our results should be interpreted as the minimum magnetic fields required within the adopted magnetar powered framework.

The dominant remaining uncertainty arises from the radio emission efficiency $\eta$, which is currently constrained using the Galactic FRB 20200428D from SGR~1935+2154. Although this event provides the only direct multi-wavelength calibration available at present, its applicability to cosmological FRBs remains uncertain. Future detections of additional FRB-associated X-ray or gamma-ray bursts will be crucial for establishing a more representative prior on $\eta$ and reducing the uncertainty in the inferred magnetic fields.

Compared with traditional spin-down methods, which require measurable periods and period derivatives and are therefore inaccessible for extragalactic FRB sources, our approach provides an alternative statistical probe of magnetar magnetic fields based on burst activity alone. While the current constraints remain limited by uncertainties in the energy conversion efficiency, the framework developed here offers a general method for connecting transient burst statistics with the physical properties of their compact-object engines.

With improved multi-wavelength observations and longer-term monitoring of repeating FRBs, the uncertainties in $\eta$ and the long-term activity evolution of magnetars can be progressively reduced. This will allow burst statistics to become an increasingly powerful diagnostic of magnetic fields in FRB sources and other magnetically powered t

\section*{acknowledgments}
This work is supported by the National Natural Science Foundation of China (grant Nos.12203013,12494575).

\bibliography{ref}
\bibliographystyle{apsrev4-2}
   
\begin{figure*}[!htb]
	\centering
	\begin{minipage}{0.25\textwidth}
		\includegraphics[width=\linewidth]{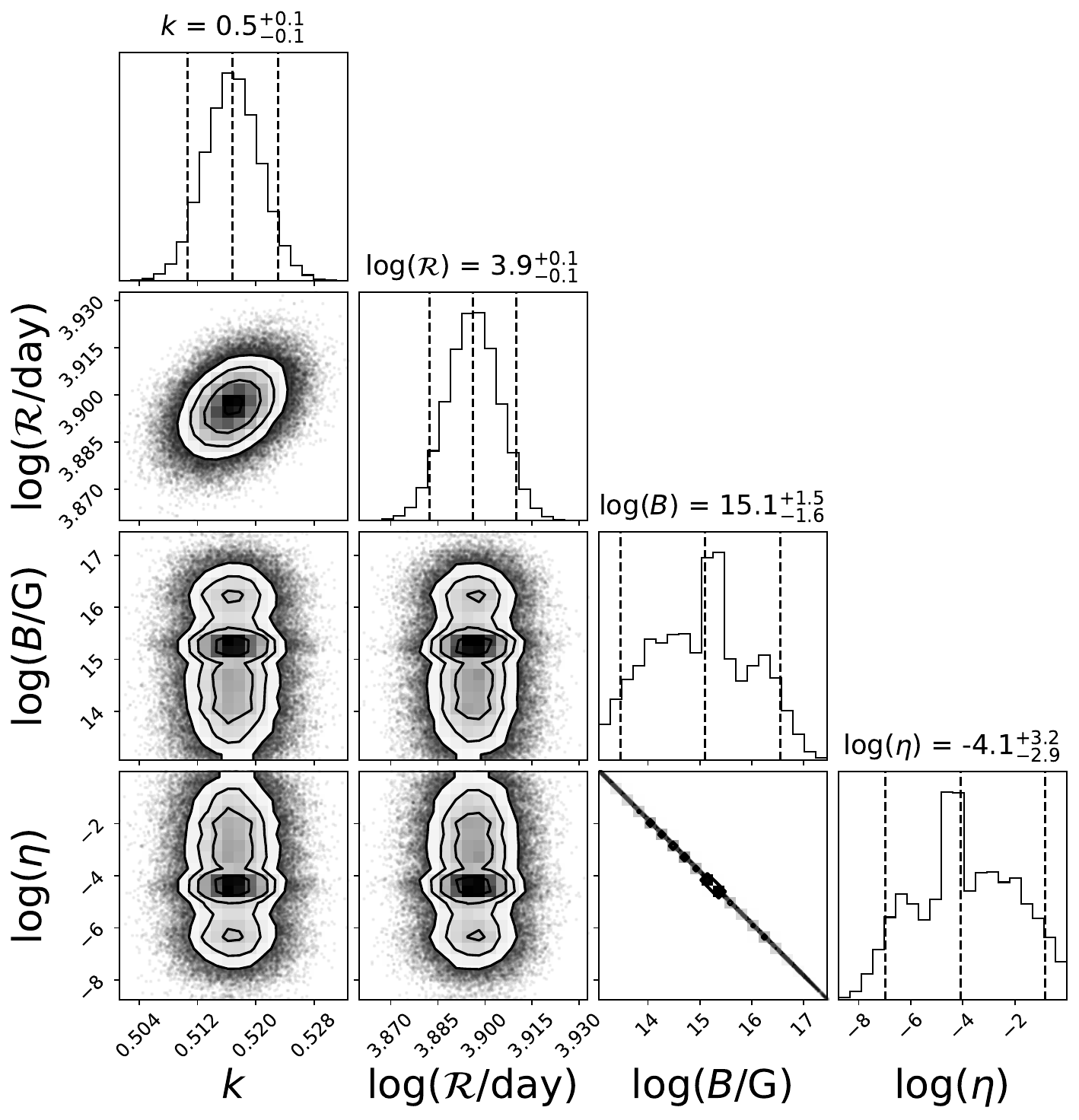}
		\subcaption{(a) FRB 20240114A}
	\end{minipage}
	\hfill
	\begin{minipage}{0.25\textwidth}
		\includegraphics[width=\linewidth]{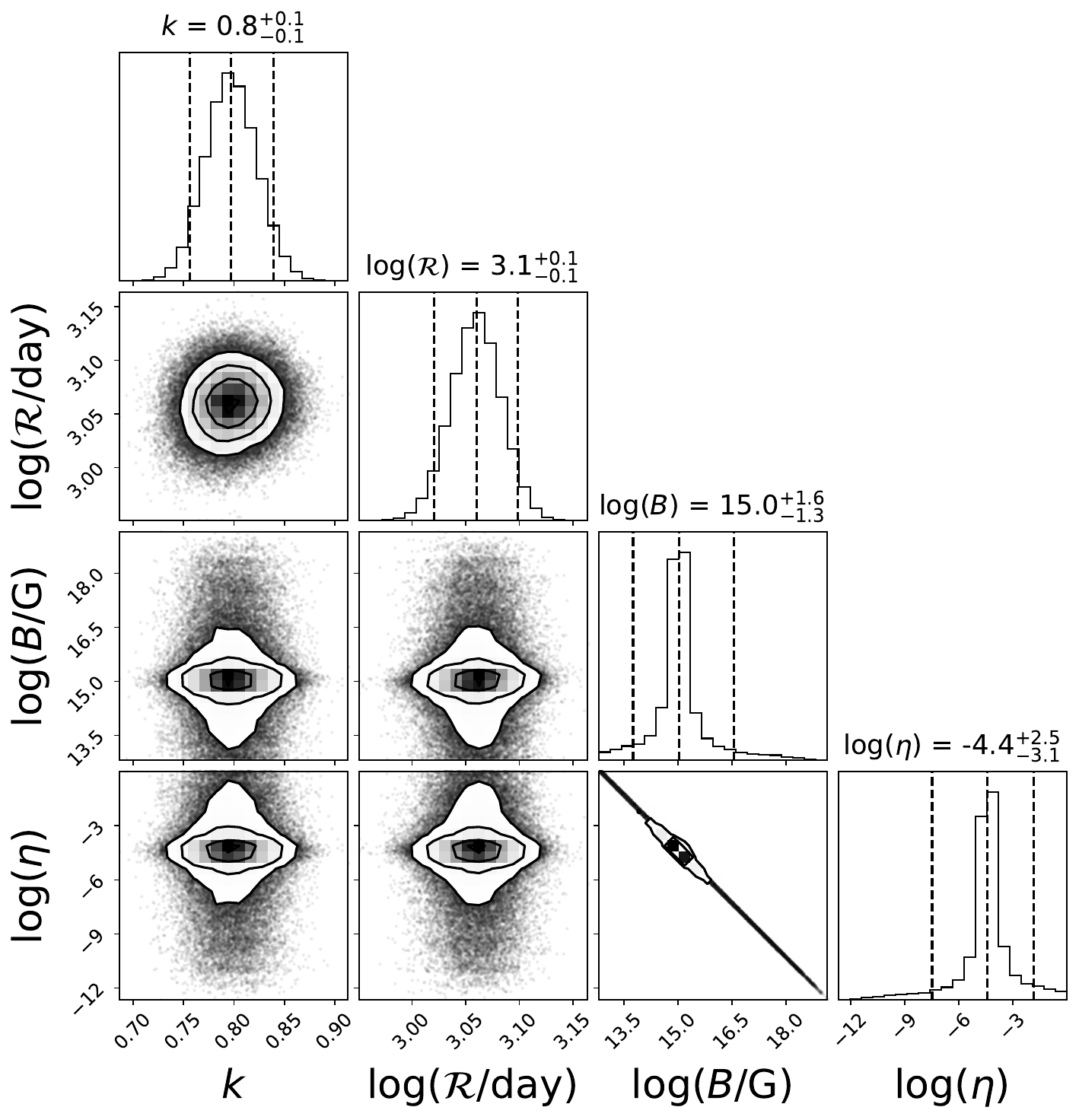}
		\subcaption{(b) FRB 20230607A}
	\end{minipage}
	\hfill
	\begin{minipage}{0.25\textwidth}
		\includegraphics[width=\linewidth]{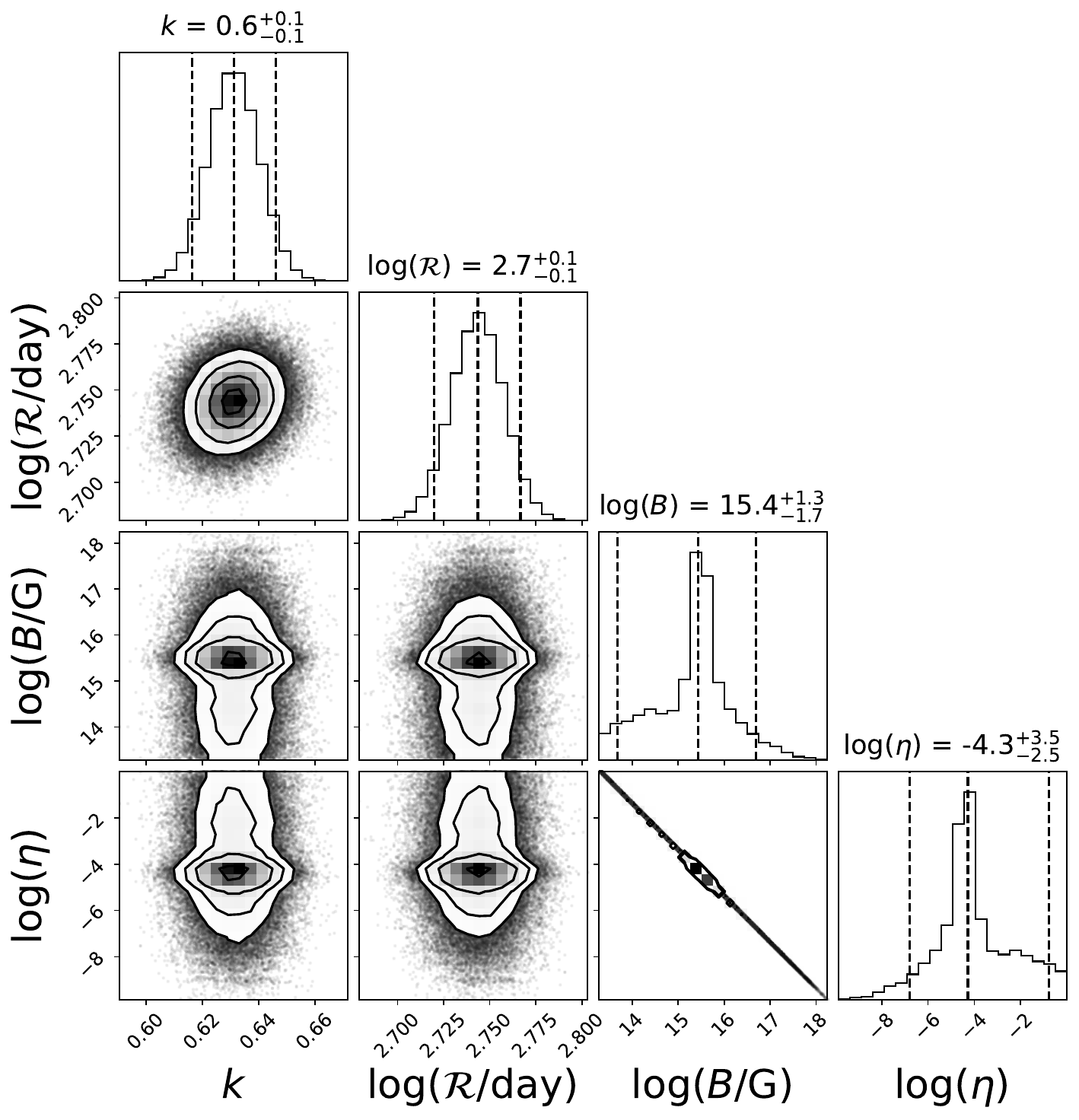}
		\subcaption{(c) FRB 20121102A}
	\end{minipage}
	
	\vspace{1em}
	\begin{minipage}{0.25\textwidth}
		\includegraphics[width=\linewidth]{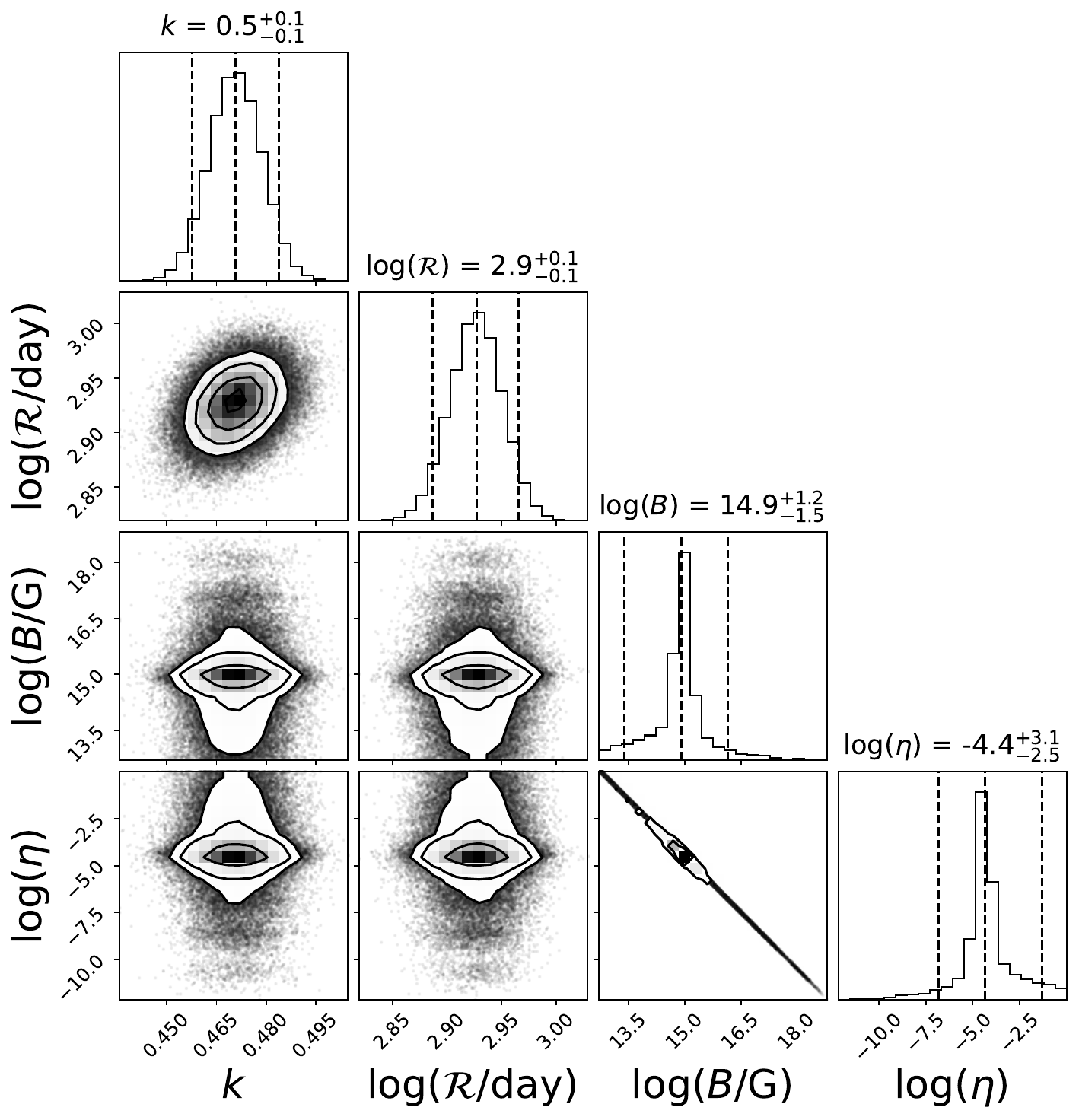}
		\subcaption{(d) FRB 20220912A}
	\end{minipage}
	\hfill
	\begin{minipage}{0.25\textwidth}
		\includegraphics[width=\linewidth]{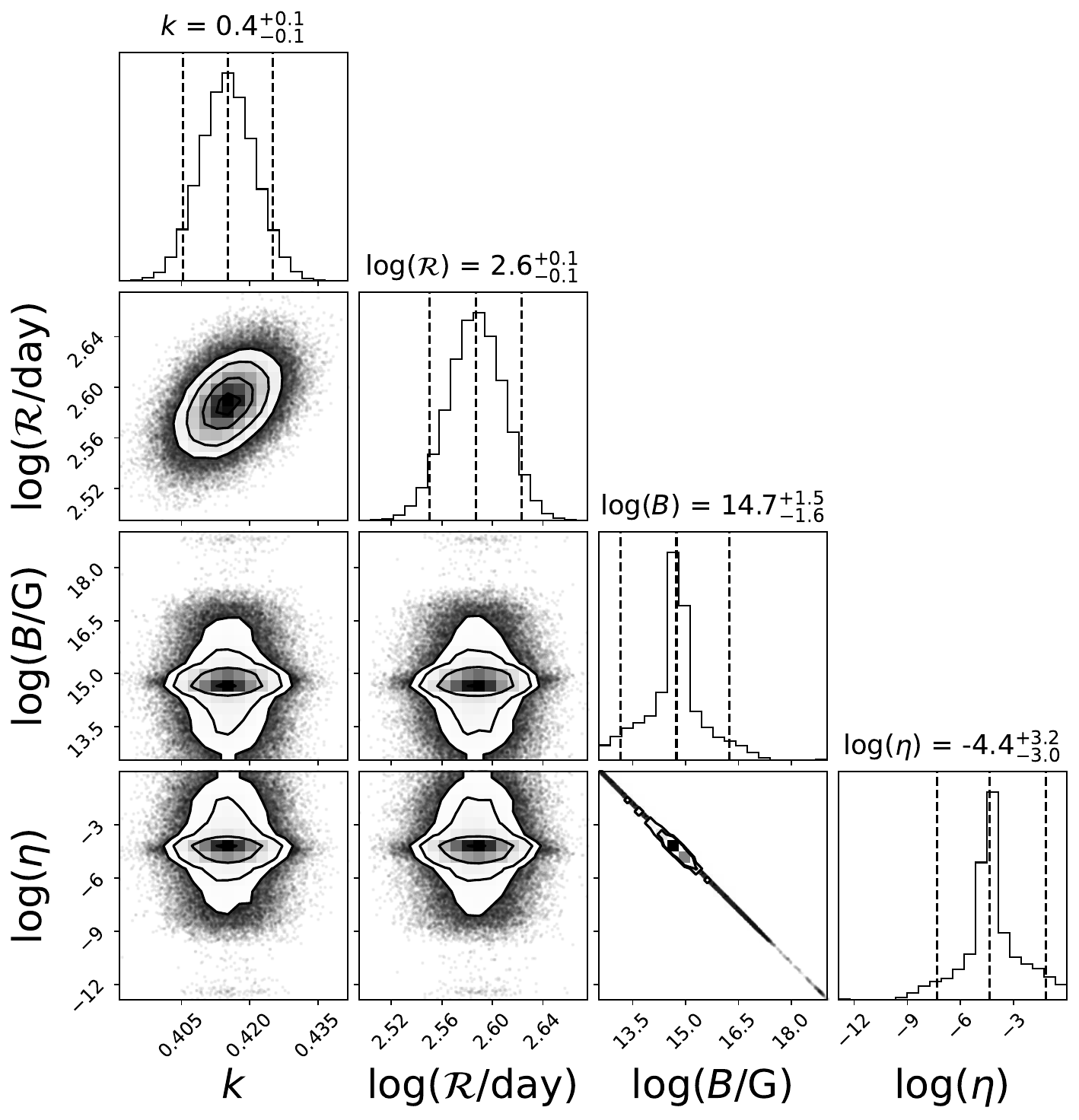}
		\subcaption{(e) FRB 20201124A}
	\end{minipage}
	\hfill
	\begin{minipage}{0.25\textwidth}
		\includegraphics[width=\linewidth]{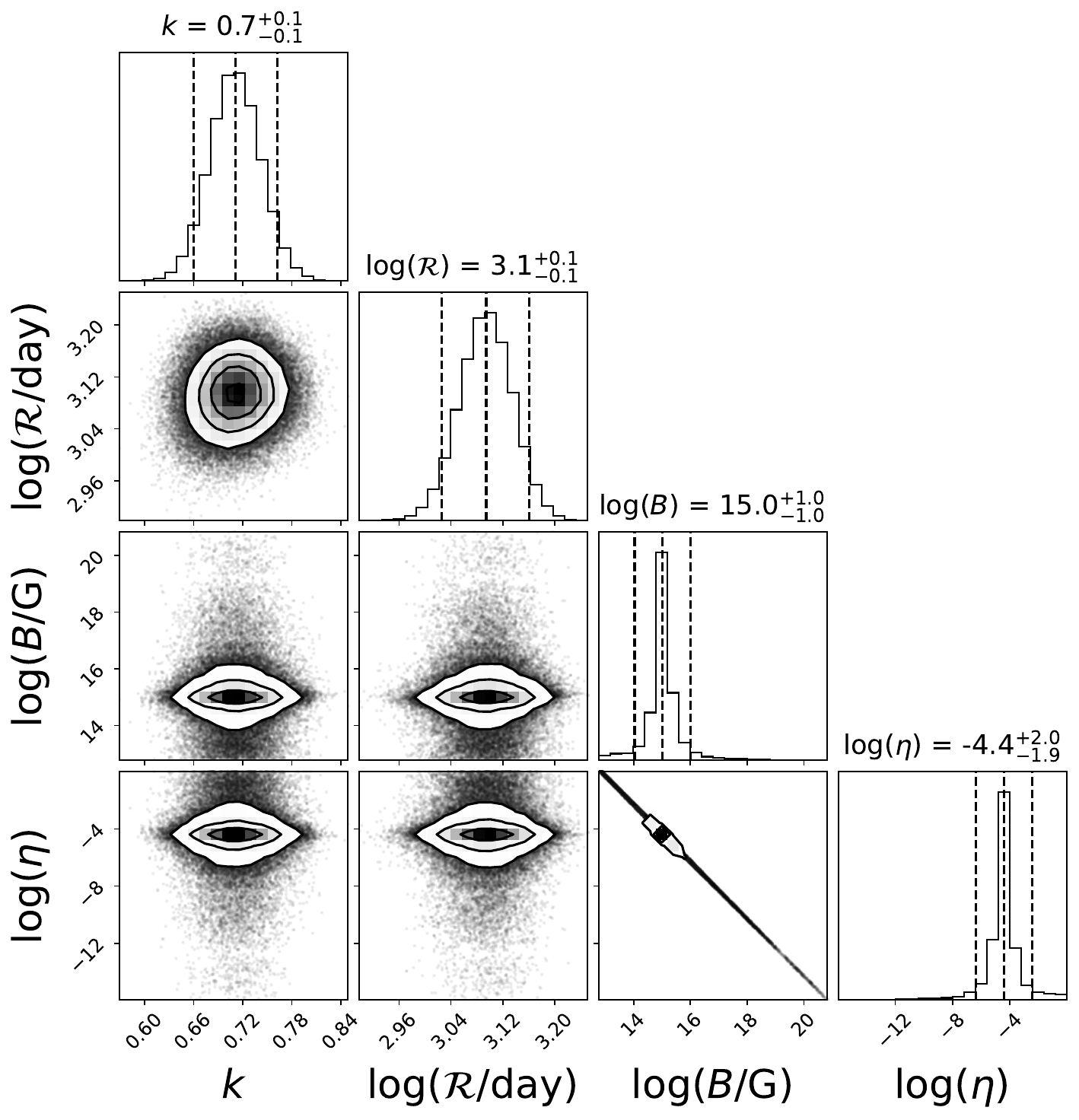}
		\subcaption{(f) FRB 20240619D}
	\end{minipage}
	
	\vspace{1em}
	\begin{minipage}{0.25\textwidth}
		\includegraphics[width=\linewidth]{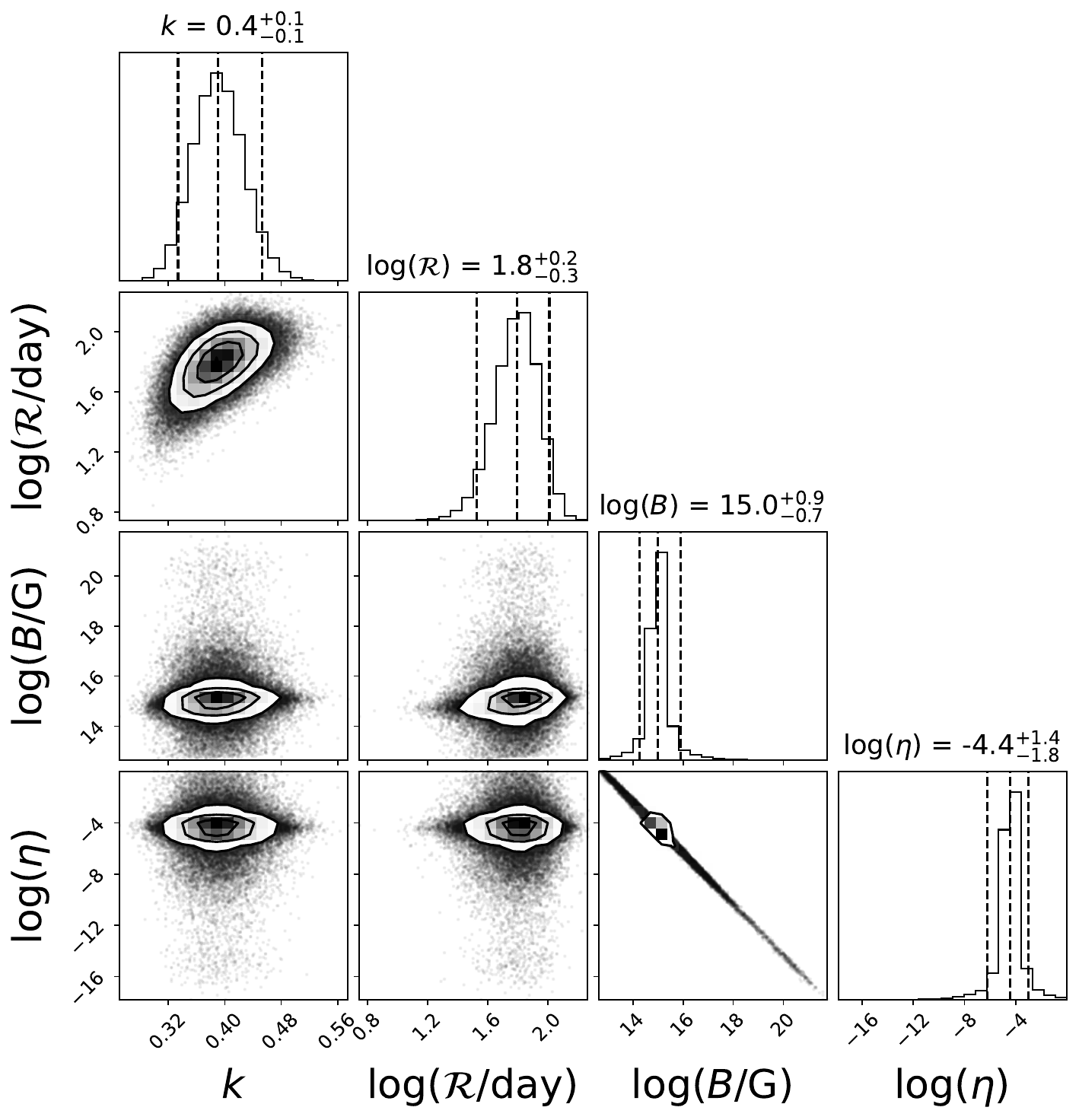}
		\subcaption{(g) FRB 20190520B}
	\end{minipage}
	\hfill
	\begin{minipage}{0.25\textwidth}
		\includegraphics[width=\linewidth]{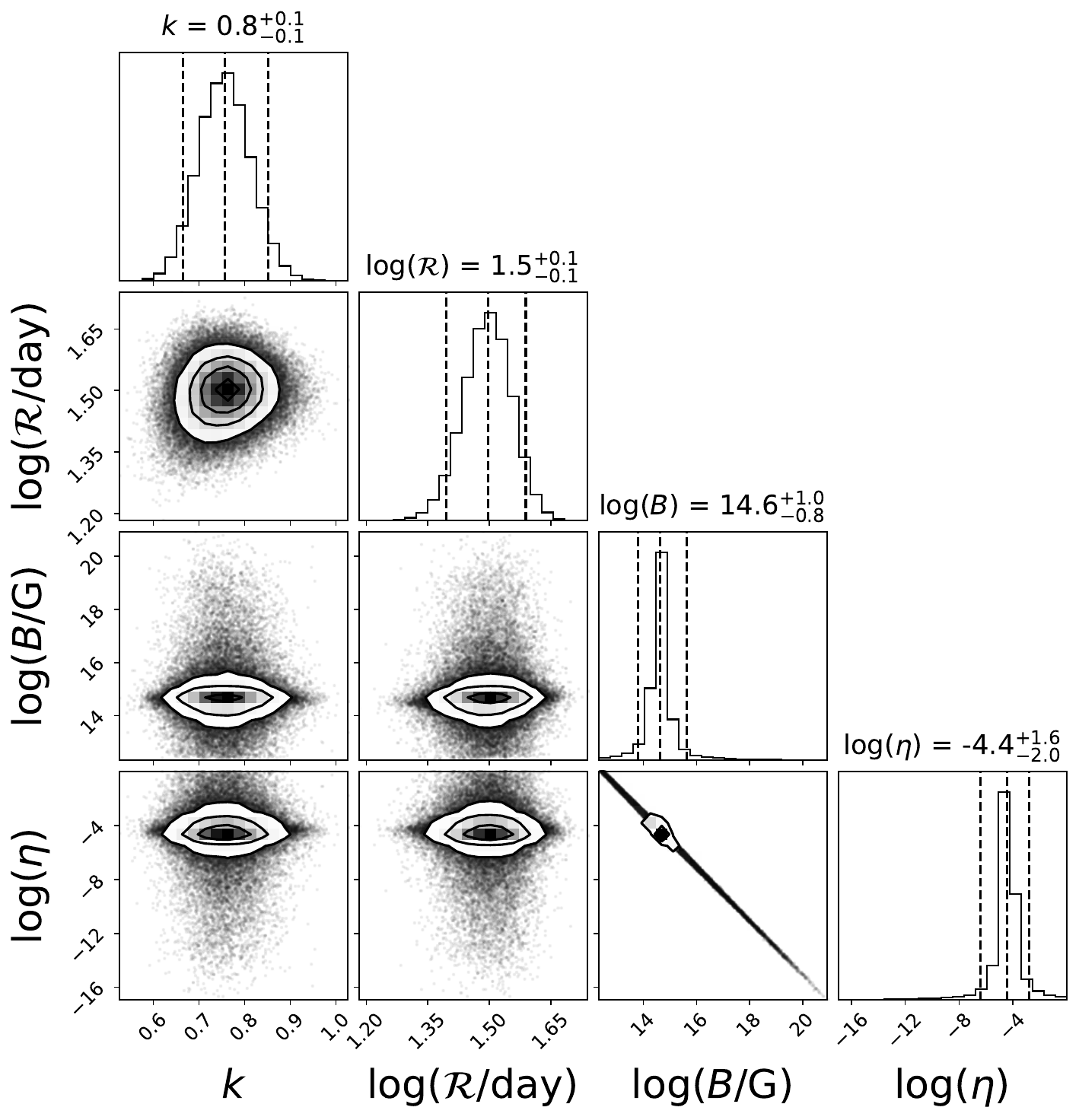}
		\subcaption{(h) FRB 20180916B}
	\end{minipage}
	\hfill
	\begin{minipage}{0.25\textwidth}
		\includegraphics[width=\linewidth]{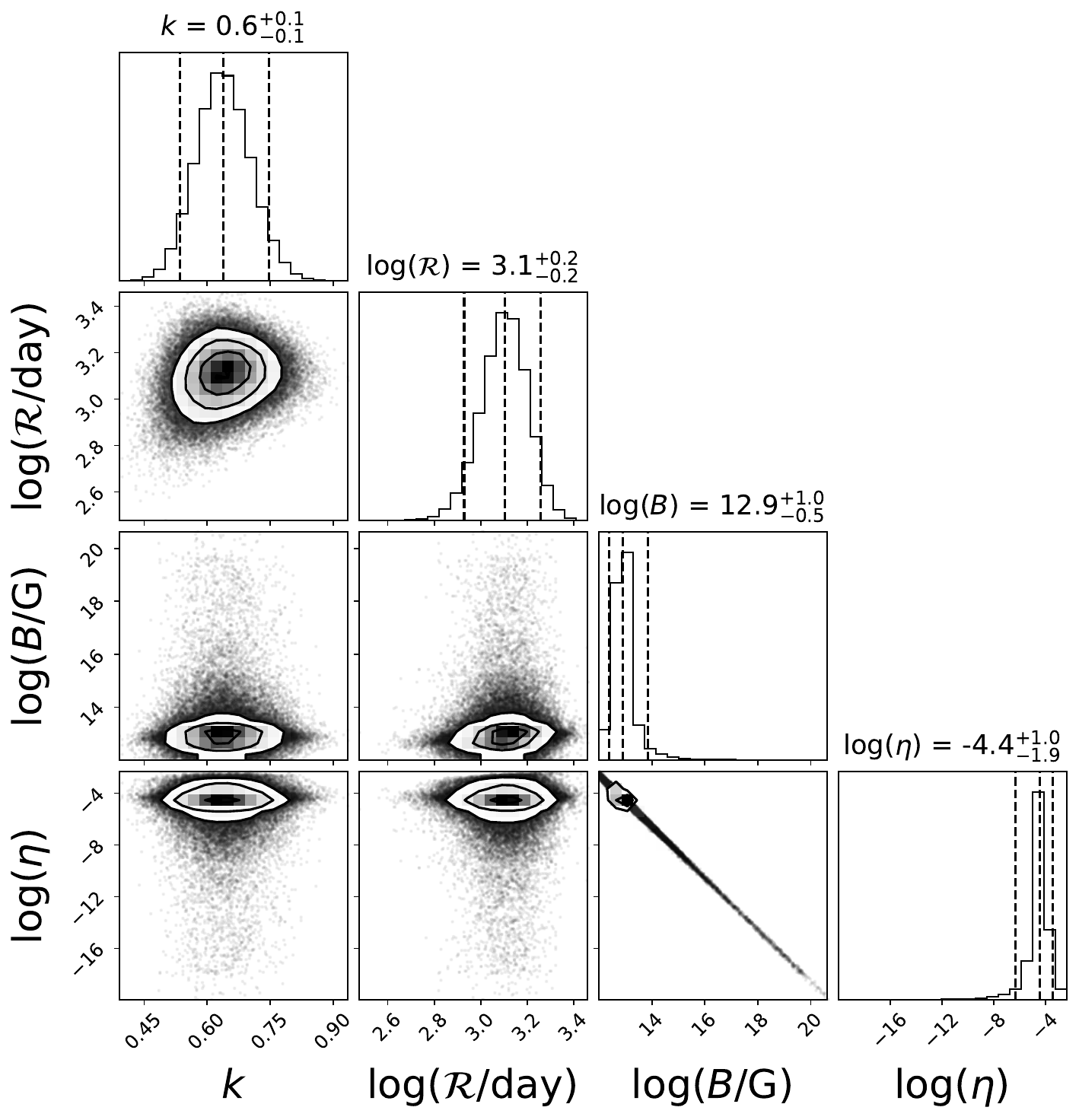}
		\subcaption{(i) FRB 20200120E}
	\end{minipage}
	
	\vspace{1em}
	\begin{minipage}{0.25\textwidth}
		\includegraphics[width=\linewidth]{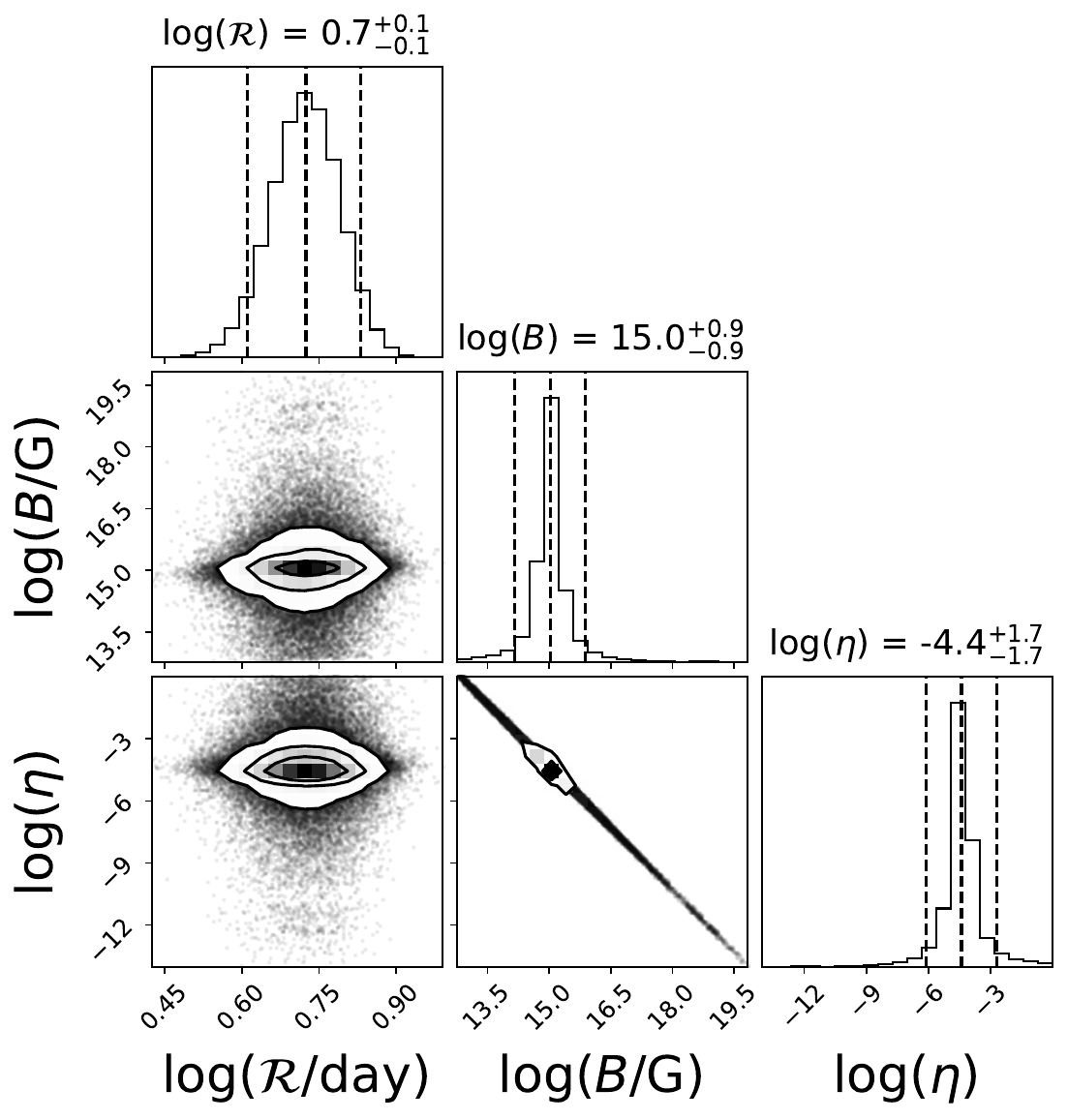}
		\subcaption{(j) FRB 20180301A}
	\end{minipage}
	\hfill
	\begin{minipage}{0.25\textwidth}
		\includegraphics[width=\linewidth]{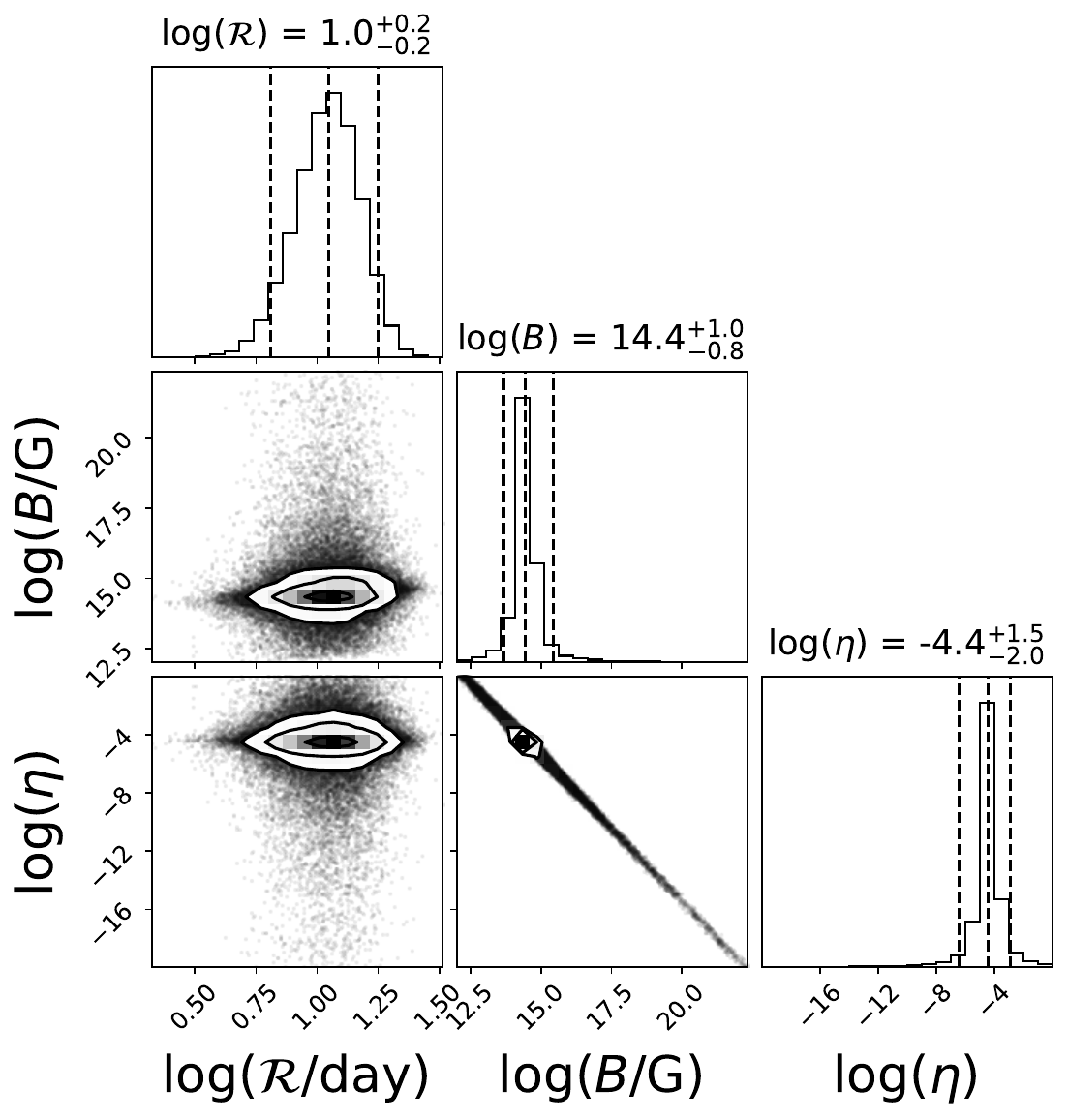}
		\subcaption{(k) FRB 20180814A}
	\end{minipage}
	\hfill
	\begin{minipage}{0.25\textwidth}
		\includegraphics[width=\linewidth]{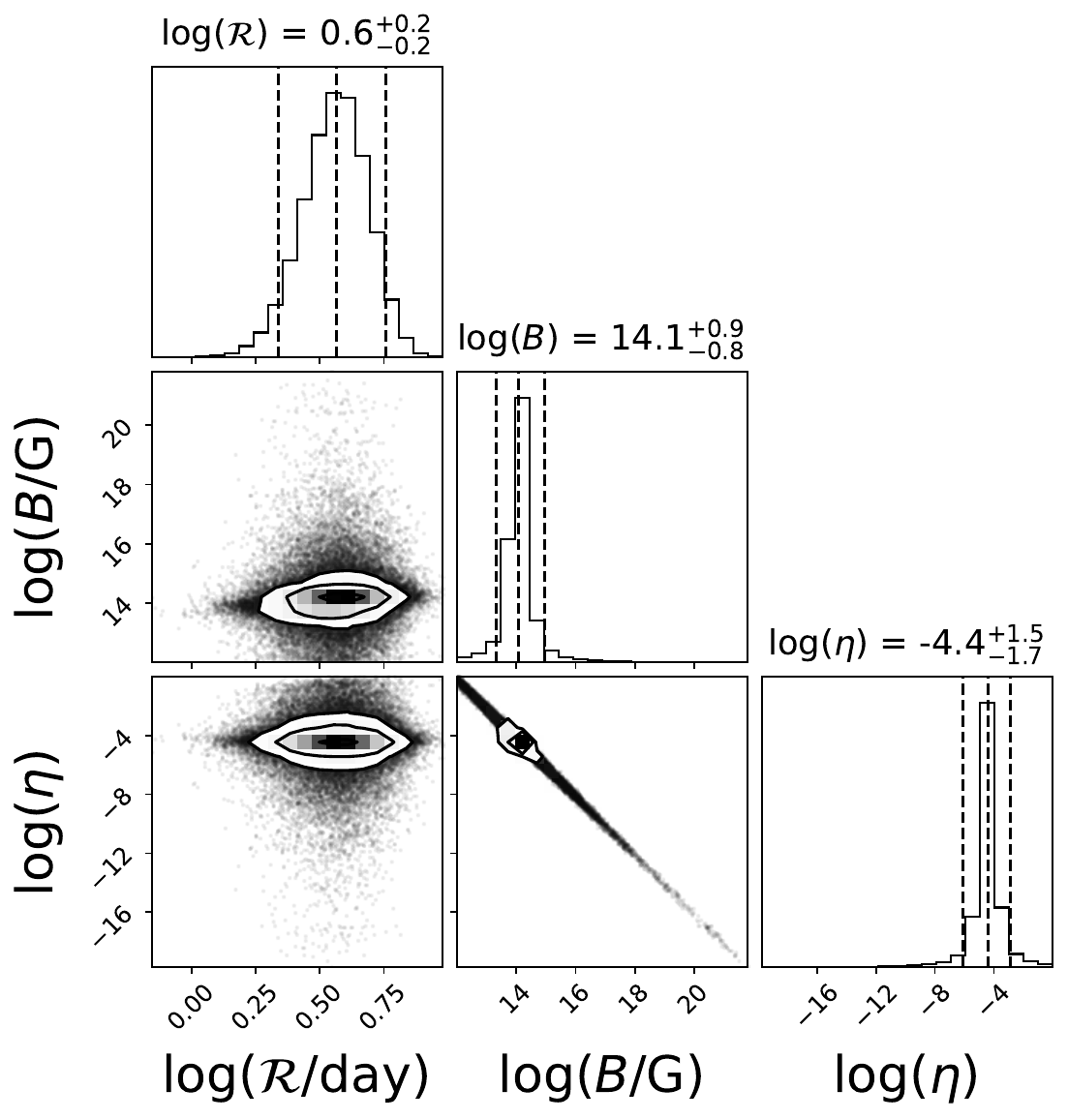}
		\subcaption{(l) FRB 20201130A}
	\end{minipage}
	
	\caption{Posterior distribution of repeating FRBs parameters under conservative estimates.}
	\label{fig:cons_4x3}
\end{figure*}

\clearpage    
\onecolumngrid
\appendix*
\section{Derivation of efficiency distribution}

For the prior distribution of $\eta$, we employ a hierarchical Bayesian approach. Set $y = \log \eta$, where we assume that $y \sim \mathrm{N}(\mu, \sigma^2)$. Based on the observed data $y_n$, the predictive distribution is given by:
\begin{equation}
	p(y|y_n) = \iint p(y|\mu, \sigma^2, y_n)p(\mu, \sigma^2|y_n)\,\mathrm{d}\mu\, \mathrm{d}\sigma^2 .
\end{equation}

\subsection{Posterior Distribution of Hyper-parameters}

For the hyper-parameters $\mu$ and $\sigma^2$, the posterior distribution is:
\begin{equation}
	p(\mu, \sigma^2|y_n) = p(y_n|\mu, \sigma^2)p(\mu, \sigma^2) .
\end{equation}

Because of the absence of prior information about the parameters, we adopt non-informative prior distributions. Assuming independence between the two parameters and uniform distributions for $\mu$ and $\log \sigma$. And set $\log \sigma = \frac{1}{2} \log {\sigma}^2$,  the Jacobian factor for the transformation from $\log \sigma$ to ${\sigma}^2$:

\begin{equation}
	\frac{\partial(\frac{1}{2}\log_{10} \sigma^2)}{\partial \sigma^2} = \frac{1}{2\sigma^2 \ln 10} ,
\end{equation}

Therefore, we can obtain:
\begin{equation}
	p(\mu, \sigma^2) \propto \sigma^{-2} .
\end{equation}

For $n$ observations, the joint posterior function becomes:
\begin{equation}  \label{eq:exponent}
	p(\mu, \sigma^2|y_n) \propto \sigma^{-n-2} \exp\left(-\frac{1}{2\sigma^2} \sum_{i=1}^n (y_{n,i} - \mu)^2\right) ,
\end{equation}

The exponential part in Eq. (\ref{eq:exponent}) can be rewritten as:
\begin{equation}
	\sum_{i=1}^n (y_{n,i} - \mu)^2 = \sum_{i=1}^n (y_{n,i} - \bar{y}_n + \bar{y}_n - \mu)^2 ,
\end{equation}

Expanding the squared term:
\begin{equation}
	\sum_{i=1}^n (y_{n,i} - \mu)^2 = \sum_{i=1}^n \left[ (y_{n,i} - \bar{y}_n)^2 + 2(y_{n,i} - \bar{y}_n)(\bar{y}_n - \mu) + (\bar{y}_n - \mu)^2 \right] ,
\end{equation}
where $\bar{y}_n = \frac{1}{n}\sum_{i=1}^n y_{n,i}$ is the sample mean and $s^2 = \frac{1}{n-1}\sum_{i=1}^n(y_{n,i} - \bar{y}_n)^2$ is the sample variance. 
Since $\sum_{i=1}^n (y_{n,i} - \bar{y}_n) = 0$ by definition of the sample mean, the cross term vanishes:
\begin{equation}
	\begin{aligned}
		\sum_{i=1}^n (y_{n,i} - \mu)^2 &= \sum_{i=1}^n (y_{n,i} - \bar{y}_n)^2 + n(\bar{y}_n - \mu)^2 \\
		& = (n-1)s^2 + n(\bar{y}_n - \mu)^2,
	\end{aligned}
\end{equation}

Therefore,  Eq. (\ref{eq:exponent}) can be rewritten as:
\begin{equation}  
	p(\mu, \sigma^2|y_n) \propto  \sigma^{-n-2} \exp\left(-\frac{1}{2\sigma^2} \left[ (n-1)s^2 + n(\bar{y}_n - \mu)^2 \right] \right).
\end{equation}

\subsection{Marginal Posterior Distributions}

\subsubsection{Conditional Posterior Distribution of $\mu$}

From the joint posterior distribution, we can extract the conditional posterior distribution of $\mu$ given $\sigma^2$. The terms involving $\mu$ in the joint posterior are:
\begin{equation}
	p(\mu|\sigma^2, y_n) \propto \exp\left(-\frac{n}{2\sigma^2}(\mu - \bar{y}_n)^2\right),
\end{equation}

This is a standard normal distribution:
\begin{equation}
	P(\mu|\sigma^2, y_n) \sim \mathrm{N}(\bar{y}_n, \sigma^2/n).
\end{equation}

\subsubsection{Marginal Posterior Distribution of $\sigma^2$}

To obtain the marginal posterior distribution of $\sigma^2$, we integrate over $\mu$:
\begin{equation}
	p(\sigma^2|y_n) \propto \int_{-\infty}^{\infty} \sigma^{-n-2}\exp\left(-\frac{1}{2\sigma^2}[(n-1)s^2 + n(\bar{y}_n - \mu)^2]\right)\mathrm{d}\mu ,
\end{equation}

Separating the terms:
\begin{equation}
	p(\sigma^2|y_n) \propto \sigma^{-n-2} \exp\left(-\frac{(n-1)s^2}{2\sigma^2}\right) \int_{-\infty}^{\infty} \exp\left(-\frac{n(\bar{y}_n - \mu)^2}{2\sigma^2}\right)\mathrm{d}\mu,
\end{equation}

The integral is a standard Gaussian integral. Using the substitution $u = \mu - \bar{y}_n$:
\begin{equation}
	\int_{-\infty}^{\infty} \exp\left(-\frac{nu^2}{2\sigma^2}\right)\mathrm{d}u = \sqrt{\frac{2\pi\sigma^2}{n}},
\end{equation}

Therefore:
\begin{equation} \label{eq:IG}
	p(\sigma^2|y_n) \propto {(\sigma^2)}^{-(n+1)/2} \exp\left(-\frac{(n-1)s^2}{2\sigma^2}\right),
\end{equation}

This is recognized as an inverse-gamma distribution:
\begin{equation}
	P(\sigma^2|y_n) \sim \text{Inverse-Gamma}\left(\frac{n-1}{2}, \frac{(n-1)s^2}{2}\right),
\end{equation}

Equivalently, this can be expressed as a scaled inverse-$\chi^2$ distribution:
\begin{equation}
	P(\sigma^2|y_n) \sim \text{Inv-}\chi^2(n-1, s^2).
\end{equation}

\subsection{Predictive Distribution}

The predictive distribution for a new observation $y$ is obtained by marginalizing over the posterior distributions of the parameters:
\begin{equation}
	p(y|y_n) = \iint p(y|\mu, \sigma^2)p(\mu|\sigma^2, y_n)p(\sigma^2|y_n)\,\mathrm{d}\mu\, \mathrm{d}\sigma^2 .
\end{equation}

\subsubsection{Integration over $\mu$}

First, we integrate over $\mu$:
\begin{equation}
	p(y|\sigma^2, y_n) = \int p(y|\mu, \sigma^2)p(\mu|\sigma^2, y_n)\mathrm{d}\mu,
\end{equation}

This is the convolution of two normal distributions: $P(y|\mu, \sigma^2) \sim \mathrm{N}(\mu, \sigma^2)$ and $P(\mu|\sigma^2, y_n) \sim \mathrm{N}(\bar{y}_n, \sigma^2/n)$.
The result is:
\begin{equation} \label{eq:nor}
	P(y|\sigma^2, y_n) \sim \mathrm{N}\left(\bar{y}_n, \sigma^2\left(1 + \frac{1}{n}\right)\right).
\end{equation}

\subsubsection{Integration over $\sigma^2$}

Next, we integrate over $\sigma^2$:
\begin{equation}
	p(y|y_n) = \int p(y|\sigma^2, y_n)p(\sigma^2|y_n)\mathrm{d}\sigma^2,
\end{equation}

Substitute Eq. (\ref{eq:IG}) and (\ref{eq:nor}) into the above expression:

\begin{equation} \label{eq:integ}
	p(y|y_n)=\int_0^\infty\frac{1}{\sqrt{2\pi\sigma^2(1+\frac{1}{n})}}\exp\left(-\frac{(y-\bar{y}_n)^2}{2\sigma^2(1+\frac{1}{n})}\right)(\sigma^2)^{-\frac{n+1}{2}}\exp\left(-\frac{(n-1)s^2}{2\sigma^2}\right)d\sigma^2,
\end{equation}

Combine the powers of ${\sigma}^2$:
\begin{equation}
	(\sigma^2)^{-\frac{1}{2}}\cdot(\sigma^2)^{-\frac{n+1}{2}}=(\sigma^2)^{-\frac{n+2}{2}},
\end{equation}

Combine the exponential terms:
\begin{equation}
	\exp\left(-\frac{(y-\bar{y}_n)^2}{2\sigma^2(1+\frac{1}{n})}\right)\exp\left(-\frac{(n-1)s^2}{2\sigma^2}\right)=\exp\left(-\frac{1}{2\sigma^2}\left[\frac{(y-\bar{y}_n)^2}{1+\frac{1}{n}}+(n-1)s^2\right]\right),
\end{equation}

Set $A=(y-\bar{y}_n)^2/(1+1/n)+(n-1)s^2$, the Eq. (\ref{eq:integ}) can be rewritten:
\begin{equation} \label{eq:b25}
	p(y|y_n)=\frac{1}{\sqrt{2\pi(1+\frac{1}{n})}}\int_0^\infty(\sigma^2)^{-\frac{n+2}{2}}\exp\left(-\frac{A}{2\sigma^2}\right)d\sigma^2,
\end{equation}

Let $u =1/\sigma^2$, then $d\sigma^2 = -1/u^2du$, the Eq. (\ref{eq:b25}) can be rewritten:
\begin{equation}
	p(y|y_n) = \frac{1}{\sqrt{2\pi(1+\frac{1}{n})}} \int_0^\infty u^{\frac{n-2}{2}} \exp\left(-\frac{Au}{2}\right)du,
\end{equation}

Gamma function integral formula:
\begin{equation}
	\int_0^\infty t^{\alpha-1}e^{-\beta t}dt=\frac{\Gamma(\alpha)}{\beta^\alpha},
\end{equation}
with $\alpha = n/2$ and $\beta = A/2$. The result of the integration is
\begin{eqnarray}
	p(y|y_n) &=& \frac{1}{\sqrt{2\pi(1+\frac{1}{n})}} \int_0^\infty u^{\frac{n-2}{2}} \exp\left(-\frac{Au}{2}\right)du \nonumber \\
	&=& \frac{1}{\sqrt{2\pi(1+\frac{1}{n})}} \frac{\Gamma\left(\frac{n}{2}\right) \cdot 2^{\frac{n}{2}}}{A^{\frac{n}{2}}},
\end{eqnarray}

Substituting the expression of $A$ into the above Equation, we can get:
\begin{eqnarray}
	p(y|y_n)&=&\frac{\Gamma\left(\frac{n}{2}\right)\cdot2^{\frac{n}{2}}}{\sqrt{2\pi(1+\frac{1}{n})}\cdot(n-1)^{\frac{n}{2}}\cdot s{}^n}\left[1+\frac{(y-\bar{y}_n)^2}{(n-1)s^2(1+\frac{1}{n})}\right]^{-\frac{n}{2}}\nonumber \\
	&\propto&\left[1+\frac{(y-\bar{y}_n)^2}{(n-1)s^2(1+\frac{1}{n})}\right]^{-\frac{n}{2}},
\end{eqnarray}

This is exactly the core form of a t-distribution with $n-1$ degrees of freedom, location parameter $\bar{y}_n$, and scale parameter $s^2\left(1+\frac{1}{n}\right)$:
\begin{equation}
	P(y|y_n) \sim t_{n-1}\left(\bar{y}_n, s^2\left(1 + \frac{1}{n}\right)\right),
\end{equation}

The factor $(1 + 1/n)$ accounts for both the sampling uncertainty in the data and the additional uncertainty arising from the estimation of the population mean $\mu$ from a finite sample of size $n$. Equivalently, this can be expressed in standardized form as:
\begin{eqnarray}
	\frac{y - \bar{y}_n}{s\sqrt{1 + \frac{1}{n}}} \sim t_{n-1}.
\end{eqnarray}

\end{document}